\documentclass{iau} 
\usepackage{graphicx}
\def\lapprox{\hbox{\lower .8ex\hbox{$\,\buildrel < \over\sim\,$}}}
\def\gapprox{\hbox{\lower .8ex\hbox{$\,\buildrel > \over\sim\,$}}}

\title[White Dwarfs and Axions] 
{White Dwarfs as Advanced Physics Laboratories. The Axion case}

\author[Jordi Isern]   
{Jordi Isern$^{1,2}$
 }

\affiliation{$^1$Institut de Ci\'encies de l'Espai (ICE,CSIC), \\ c/Can Magrans, Campus UAB,
08193 Cerdanyola, Spain \\ email: {\tt isern@ice.cat} \\[\affilskip]
$^2$Institut d'Estudis Espacials de Catalunya (IEEC), \\ Edifici Nexus 201,
c/Gran Capit\`a 2-4, 08034 Barcelona, Spain \\email: {\tt isern@ieec.cat}}

\pubyear{2019}
\volume{357}  
\jname{White Dwarfs as Probes of Fundamental Physics and Tracers of Planetary, Stellar, and Galactic Evolution}
\editors{M. Barstow, S. Kleinman, J. Proven\c cal, \& L. Ferrario eds.}
\begin{document}

\maketitle

\begin{abstract}
The shape of the luminosity function of white dwarfs (WDLF) is sensitive to the characteristic cooling time and, therefore, it can be used to test the existence of additional sources or sinks of energy such as those predicted by alternative physical theories. However, because of the degeneracy between the physical properties of white dwarfs and the properties of the Galaxy, the star formation history (SFH) and the IMF, it is almost always possible to explain any anomaly as an artifact introduced by the star formation rate. To circumvent this problem there are at least two possibilities, the analysis of the WDLF in populations with different stories, like disc and halo, and the search of effects not correlated with the SFH. These procedures are illustrated with the case of axions.
\keywords{white dwarfs, axions.}
\end{abstract}

\firstsection 
\section{Introduction}
Often non-standard theories predict the existence of particles which existence and properties cannot be tested in the terrestrial laboratories as a consequence of the large energies involved. One way to alleviate the problem is the use of stars to constrain their properties, either looking for a decay or a change of these properties during the trip, or examining the perturbations they introduce into the \emph{normal} evolution of stars 
(\cite[Raffelt, 1996]{raff96}).

 Because the evolution of white dwarfs is a relatively simple process of cooling, the basic ingredients necessary to predict their behavior are well identified, and there is a solid observational background to test the theoretical results, these stars have proved to be excellent laboratories for testing new ideas in Physics (\cite[Isern \& Garc\'{\i}a-Berro, 2008]{iser08a}). This procedure has allowed to put bounds on the mass of axions (\cite[Raffelt, 1986]{raff86}; \cite[Isern \etal, 1992]{iser92}; \cite[Isern \etal, 2008]{iser08b}), on the neutrino magnetic momentum (\cite[Blinnikov \& Dunina-Barkovskaya, 1994]{blin94}), the secular drift of the Newton gravitational constant (\cite[Vila, 1976]{vila76}; \cite[Garc\'{\i}a-Berro. E., \etal, 1995]{garcb95}), the density of magnetic monopoles (\cite[Freese, 1984]{free84}) and WIMPS (\cite [Bertone \& Fairbairn, 2008]{bert08}), as well as constraints on properties of extra dimensions (\cite[Malec, \& Besiada, 2001]{male01}), on dark forces (\cite[Dreiner \etal, 2013]{drei13}), on modified gravity (\cite[Saltas \etal, 2019]{salt19}), and formation of black holes by high energy collisions (\cite[Giddings \& Mangano, 2008]{gidd08}). In this talk only axions will be discussed.

According to the standard theory, white dwarfs are the last stage of the evolution of low and intermediate mass stars. Since their core is almost completely degenerate, they cannot obtain energy from nuclear reactions and their evolution is just a process of contraction and cooling\footnote{See \cite[Iben \& Tutukov (1984)]{iben84}, \cite[Koester \& Shoenberner (1986)]{koes86}, \cite[D'Antona \& Mazzitelli (1989)]{dant89}, \cite[Isern \etal,  (1998)]{iser98}, \cite[Fontaine \etal,  (2001)]{font01}, \cite[Hansen (2004)]{hans04}, \cite[Althaus \etal, (2010)]{alth10}, \cite[Garc\'{\i}a-Berro \& Oswalt ( 2016)]{garcb16} for detailed descriptions of the cooling process.}. The main sources of energy are the gravo-thermal readjustment of the structure, represented by the first two terms of the r.h.s. of Equation~\ref{eqe}, the gravitational settling of heavy species like $^{22}$Ne, $g_s$, the latent heat and sedimentation associated to crystallization, times the crystallization rate $\dot m_s$ and any other exotic source or sink of energy ($\dot\varepsilon_e$). The l.h.s. of Equation~\ref{eqe} contains the sinks of energy, photons and neutrinos. This equation has to be complemented with a relationship connecting the temperature of the core with the luminosity of the star. Typically $L \propto T_c^\alpha$ with $\alpha \approx 2.5 - 2.7$.

\begin{equation}
L + {L_\nu }  =  - \int\limits_{{M_{WD}}} {{c_V}\frac{{d{T_C}}}{{dt}}dm - \int\limits_{{M_{WD}}} {T{{\left( {\frac{{\partial P}}{{\partial T}}} \right)}_{V,x}}\frac{{dV}}{{dt}}dm +
g_s + \left( {{l_s} + {e_s}} \right){{\dot m}_s} \pm \left( {{\dot\varepsilon _e}} \right)} } 
\label{eqe}
\end{equation}

\subsection{Energy losses by neutrinos}
The importance of neutrino losses in the evolution of white dwarfs was early recognized by \cite[Vila (1968)]{vila68} and \cite[Savedoff \etal, (1969)]{save69}. During the first stages of cooling, when the star is still very hot, the energy losses are dominated by the plasma and photo neutrino processes and, as soon as the temperature decreases, by neutrino bremsstrahlung  ( \cite[Iben \& Tutukov 1984]{iben84}; \cite[D'Antona \& Mazzitelli, 1989]{dant89}). Besides their role in the cooling, neutrinos force the thermal structures produced by AGB stars to converge towards a unique one, guaranteeing in this way the uniformity of models dimmer than $ \log (L/L_\odot) < -1.5$ .
However, in spite of the enormous progress experienced by the physics of neutrinos, several questions still remain. For instance, are neutrino Dirac or Majorana particles, do sterile neutrinos exist, which is their mass spectrum, do they have magnetic momentum?. This last problem, for instance, is specially important since the existence of a magnetic dipole momentum can notably enhance the neutrino losses in white dwarfs (\cite[Blinnikov \& Dunina-Barkovskaya, 1994]{blin94}).

\subsection{Influence of the DA non-DA character}
The luminosity strongly depends on the properties of the envelope (mass, chemical composition and structure) as well as on the total mass and radius of the white dwarf. The main characteristics of the envelope is its tendency to become stratified, the lightest elements tending to be placed on top of the heaviest ones as a consequence of the strong gravitational field. However, this behavior is counterbalanced by convection, molecular diffusion and other processes that tend to restore the chemical homogeneity. In any case, the $\sim 80$\% of white dwarfs shows the presence of H--lines in their spectra while the remaining $\sim 20$\% not. This proportion is not constant along the cooling sequence. The first ones are generically called DAs and the second ones non--DAs. The most common interpretation is that the DAs have a double layered envelope made of H ($M_{\rm H} \sim 10^{-4}M_{\rm WD}$) and He ($M_{\rm He} \sim 10^{-2}M_{\odot}$) while the non-DAs have just a single He layer or an extremely thin H layer. An additional complication is that the initial conditions at the moment of formation are not well known and for the moment it is not possible to disentangle which part of this behavior is inherited and which part is evolutive, although probably both are playing a role (\cite[Althaus \etal, 2010]{alth10}).
In principle, it is possible to adjust the parameters of the AGB progenitors to obtain 25\% of white dwarfs completely devoided of the hydrogen layer. But, since the relative number of DA/non-DA stars changes during their evolution, a mechanism able to transform this character must exist (\cite[Shipman 1997]{ship97}).

It is commonly accepted that DAs start as the central star of a planetary nebula and asteroseismological data suggest they are born with a hydrogen layer of a mass in the range of $10^{-8} - 10^{-4}$ M$_\odot$. As the star cools down, the outer convection zone  deepens and, depending on the mass, completely mixes the hydrogen layer with the larger helium layer in such a way that DAs turn out into non-DAs and, consequently, the ratio DA/non-DA decreases with the effective temperature.

The evolution of non-DAs is more complex. They are born as He-rich central stars of planetary nebulae and, as they cool down they look as PG 1159 stars first and DO after. The small amount of hydrogen present in the envelope floats up to the surface and when the temperature is $\sim 50,000$ K forms an outer layer thick enough to hide the helium layer to the point that the star becomes a DA. When the temperature goes below 30,000 K, the convective helium layer engulfs the hydrogen one and the white dwarf recovers the non-DA character, now as a DB, and, as it continues to cool down, it becomes a DC. Notice that a fraction of DCs has a DA origin.
Besides the phenomenological differences between DA and non-DA families, the most important property is that they cool down at different rates since hydrogen is more opaque than helium.

\section{The axion case}
The so called \emph{strong CP problem}, i.e. the existence in the Lagrangian of Quantum Chromodynamics of a term, not observed in Nature, that violates the charge-parity symmetry, is one of the most important problems that has to face the Standard Model.
One possible solution consists on the introduction of a new symmetry that breaks at energies of the order of $f_a \sim 10^9-10^{11}$~GeV and gives raise to a new particle, the \emph{axion}.  This particle is a boson and has a mass $m_{\rm a} = 6\,{\rm meV}(10^9{\rm GeV}/f_{\rm a})$ that is not fixed by the theory. The larger is the mass, the larger is the interaction with 
matter~(\cite[Raffelt, 1996]{raff96}).

The interaction with photons and fermions is described as:
\begin{equation}
\mathit{L}_a  =  - \frac{1}{4}g_{a\gamma } F_{\mu \nu } \tilde F^{\mu \nu }  - \sum\limits_{fermions} {g_{ai} a\overline {\psi _i } \gamma _5 \psi _i } 
\end{equation}
\noindent
where $g_{a\gamma \gamma}({\rm GeV}^{-1})$ and  $g_{ai}$ are the corresponding coupling constants,
$F$ is the electromagnetic field tensor and $\gamma_5$ is the corresponding
 Dirac field. The values that these constants take are model dependent.
 
 There are several models of axions (\cite[Dias \etal, 2014]{dias14}). Here only the DFSZ one is considered (\cite[Dine, Fishler \& Srednick, 1981]{dine81}; \cite[Zhimitskii, 1980]{zhim80}) since it predicts a profuse production of axions in the hot and dense interior of white dwarfs as a consequence of the interaction  with electrons.  In this model, the adimensional axion-electron coupling constant is related to the mass of the axion through
 \begin{equation}
 g_{\rm ae} = 2.8\times 10^{-14} m_{\rm a}[{\rm meV}]\cos^2\beta
 \label{gae}
 \end{equation}
 where $\cos^2 \beta$ is a free parameter. For white dwarfs in the luminosity range  $8.5 \le M_{\rm bol}  \le 12.5$ the production of neutrinos  and axions is dominated by bremsstrahlung.
 
 The emission of thermal axions is similar to that of thermal neutrinos\footnote{With the exception of the Primakoff effect.}. There are, however, subtle differences introduced by the boson character of axions. One is that when the white dwarf cools down, neutrino emission, $\dot \epsilon_{\rm bremss} \propto T^7$ (\cite[Itoh \etal, 1996]{itoh96}), is quenched before axionic emission, $\dot \epsilon_{\rm a} \propto T^4$ (\cite[Nakagawa \etal, 1987]{naka87}; \cite[Nakagawa \etal, 1988]{naka88}), as it can be seen in Fig.~\ref{fig1}\footnote{\cite[Altherr \etal, (1994)]{althe94} have claimed that these values were understimated by a factor $\sim 4$. If this were correct this would mean that $g_{ae}$ should be divided by a factor $\sim 2$}. 
The other effect is that when white dwarfs are hot, axion emission modifies the temperature profile and reduces the neutrino losses (\cite[Miller Bertolami \etal, 2014]{mill14}).

\section{Variable white dwarfs}
During the process of cooling, white dwarfs cross some regions of the H-R diagram where they become unstable and pulsate. The multifrequency character and the size of the period of pulsation ($10^2-10^3$~s)
indicate that they are g-mode pulsators\footnote{Radial pulsations have shorter periods, $P\sim 10$~s.}. As a variable white dwarf cools down, the oscillation period, $P$, changes as a consequence of the changes in the thermal and mechanical structure. This secular drift can be approximated by (\cite[Winget \etal, 1983]{wing83}):
\begin{equation}
\frac{{\dot P}}{P} =  - a\frac{{\dot T}}{T} + b\frac{{\dot R}}{R}
\label{pdot}
\end{equation}
where $a$ and $b$ are positive constants of the order of unity. The first term of the r.h.s. reflects the decrease of the Brunt-V\"ais\"al\"a frequency with the temperature, while the second term reflects the increase of the frequency induced by the residual gravitational contraction. 

There are at least three types of variable white dwarfs, the DOV, DBV, and DAV. In the first case, gravitational contraction is still significant and the second term of Eq.~\ref{pdot} is not negligible,  for which reason $\dot P$ can be positive or negative. The DBV stars are characterized by the lack of H-layer, their effective temperatures  are in the range  of $23,000 - 30,000$~k and the secular drift is always positive and in the range of $10^{-13}-10^{-14}$~ss$^{-1}$. The DAV white dwarfs, also known as ZZ Ceti stars, are characterized by the presence of a thin atmospheric layer of pure hydrogen. Their effective temperature is in the range of $12,000-15,000$~K and their period drift, always positive, is of the order of $\sim 10^{-15}$~ss$^{-1}$.
Therefore, these secular drifts can be used to test the predicted evolution of white dwarfs and, if the models are reliable enough, to test any physical effect able to change the pulsation period of these stars. In order of magnitude,

\begin{equation}
\frac{{{L_{0}} + {L_x}}}{{{L_{0}}}} \approx \frac{{{{\dot P}_{obs}}}}{{{{\dot P}_{0}}}}
\end{equation}
where $\dot P_{obs}$ is the observed period drift,  $L_0$ and $\dot P_0$ are obtained from standard models and $L_x$ is the extra luminosity necessary to fit the observed period (\cite[Isern \etal, 1992]{iser92}) .

\begin{figure}[h]
\begin{minipage}{15pc}
\includegraphics[width=15pc]{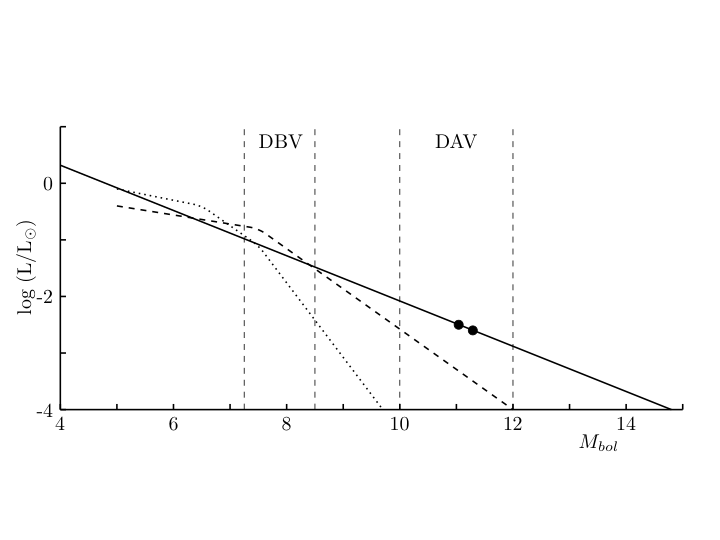}
\caption{\label{fig1} {\footnotesize Evolution of the photonic (solid), nautrinic (dotted), and
axionic (dashed) luminosities as the white dwarf fades. The vertical bands display the loci of the different families of variables. }}
\end{minipage}\hspace{2pc}%
\begin{minipage}{15pc}
\includegraphics[width=15pc,angle=90,clip=true,trim=2.5cm 0cm 0cm 0cm]{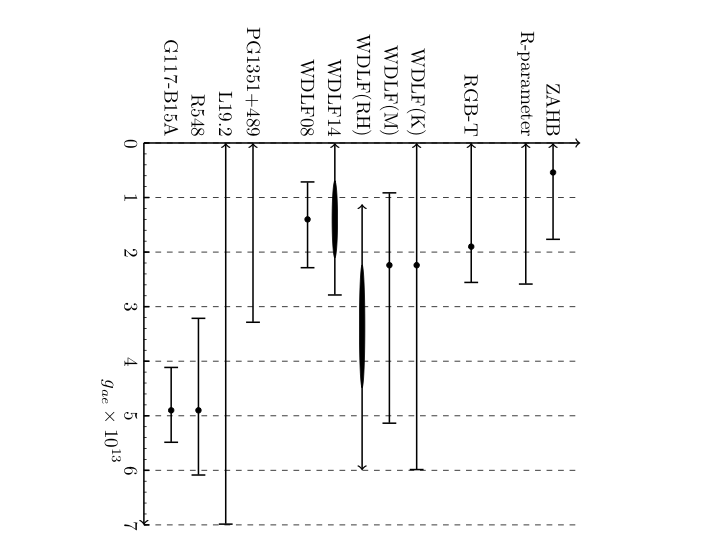}
\caption{\label{fig2} {\footnotesize Axion-electron strengths obtained from different stellar constraints.}}
\end{minipage} 
\end{figure}

G117-B15A is a ZZ Ceti star discovered by \cite[McGraw \& Robinson (1976)]{mcgr76} that has been monitored since then.  The first measurement of $\dot P$ gave  a value that was a factor two larger than expected (\cite[Kepler \etal,  1991]{kepl91}). The temperature of such star is low enough to neglect the radial term of Eq.~\ref{pdot} and the influence of neutrinos.  These two facts led to \cite[Isern \etal, (1992)]{iser92} to postulate axions of the DSKZ type  with $g_{\rm ae}\approx 2.2\times 10^{-13}$  as responsible of the anomalous cooling. The analysis of \cite[C\'orsico \etal,  (2001)]{cors01} and \cite[Bischoff-Kim \etal,  (2008)]{bisc08} showed that the presence of axions accelerates the drift but not changes the period of pulsation. The most recent estimation of the drift of the 215 s pulsation period 
is $\dot P_{\rm obs} = (4.07 \pm 0.61)\times 10^{-15}$ s\,s$^{-1}$ (\cite[Kepler \etal,  2012]{kepl12}), which suggests  $ g_{\rm ae} = 4.9\times 10^{-13}$ (Fig.~\ref{fig2}) from full model fitting (\cite[C\'orsico \etal, 2012a]{cors12a}).

Similar anomalous drifts have been found in R548, the ZZ Ceti star itself (\cite[Mukadam \etal, 2013]{muka13}), and in L19.2 (\cite[Sullivan \& Chote, 2015]{sull15}), which could be accounted introducing axions with $g_{\rm ae}\times 10^{13}=4.8$~and~$<7$ respectively (\cite[C\'orsico \etal, 2012b]{cors12b}; \cite[C\'orsico \etal, 2016 ]{cors16}).

These analysis indicate that the cooling rate is larger than expected if the pulsation modes are trapped in the outer envelope. This poses a problem since the trapping strongly depends on the detailed chemical gradients and the chemical structure of these layers, which are built during the AGB phase, is very sensitive to the methods used  to treat convection and pulses during this epoch. Other problems come from the fact that these regions are partially degenerate and not all the physical inputs, specially axion emissivities, are correctly computed in this regime. Furthermore, there are still many uncertainties in the equation of state and opacities.

In the case of DB white dwarfs the drift has been measured in PG1351+489 (\cite[Redaelli \etal, 2011]{reda11}), $\dot P_{\rm obs} = (2.0 \pm 0.9)\times 10^{-13}$ s\,s$^{-1}$, which  provides a bound of $g_{ae}\times 10^{13}<3.3$ (\cite[Battich \etal, 2016]{batt16}). See Figure~\ref{fig2}. Notice that at these temperatures, neutrinos are still active and their emission can be affected by axions or even by the existence of a hypothetical magnetic momentum. 

\section{The luminosity function}
\label{seclf}

The luminosity function (LF) is defined as the number density of white dwarfs of a given luminosity per unit magnitude interval:
\begin{equation}
n\left( L \right) = \int\limits_{{M_l}}^{{M_u}} {\Phi \left( M \right)\Psi \left( {{T_G} - {t_{cool}} - {t_{ps}}} \right){\tau _{cool}}dM} 
\label{lf}
\end{equation}
where $M$ is the mass of the parent star (for convenience all white dwarfs are labeled with the mass of the zero age main sequence progenitor), $t_{cool}$ is the cooling time down to luminosity $L$, $\tau_{cool} = dt/dM_{bol}$ is the characteristic cooling time of the white dwarf, $t_{ps}$ is the lifetime of the progenitor of the white dwarf and $T_G$ is the age of the Galaxy or the population under study, and $M_u$ and $M_l$ are the maximum and the minimum mass of the main sequence stars able to produce a white dwarf, therefore $M_l$ satisfies the condition $T_G=t_{cool}(L,M_l)+t_{ps}(M_l)$. $\Phi(M)$ is the initial mass function and $\Psi(t)$ is the star formation rate (SFR) of the population under consideration. Additionally, hidden, there is an initial-final mass function connecting the properties of the progenitor with those of the white dwarf. In order to compare theory with observations, and since the total density of white dwarfs is not yet  well known, the computed luminosity function is usually normalized to the bin with the smaller error bar, usually $\log L/L_\odot \simeq 3$. 

This equation contains three sets of terms, the observational ones, $n(L)$, the stellar ones, $t_{\rm cool},\tau_{\rm cool}, t_{\rm PS}, M_{\rm U}, M_{\rm i}$, plus the initial final mass function 
(\cite[Catalan \etal, 2008]{cata08}), and the galactic ones $\Phi$ and $\Psi$.

\begin{figure}[h]
\begin{minipage}{15pc}
\includegraphics[width=15pc,clip=true, trim= 0cm  2.5cm 0cm 0.5cm]{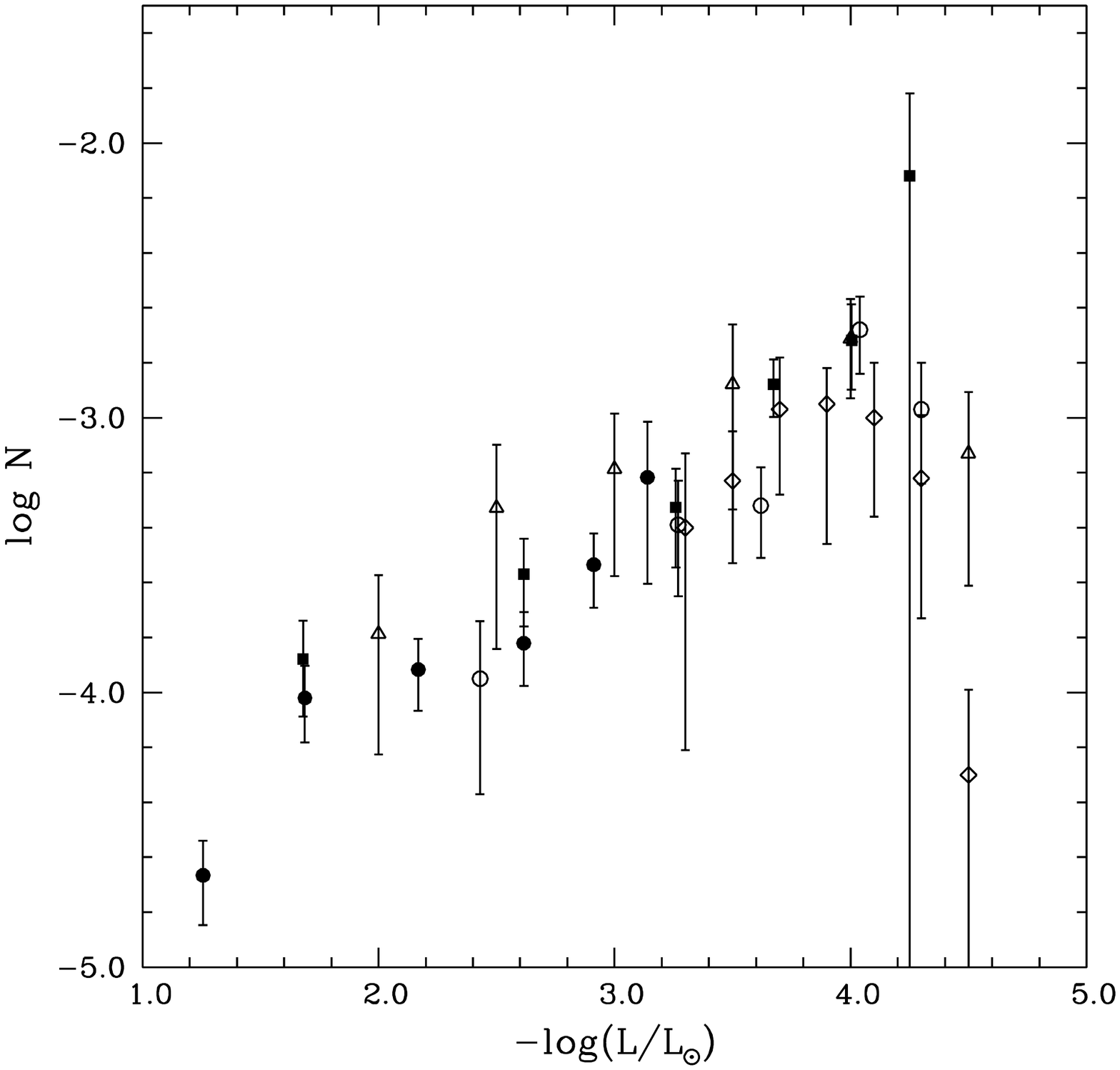}
\caption{\label{fig3} {\footnotesize Luminosity functions obtained before the large cosmological surveys -- \cite[Liebert, Dahn \& Monet (1988)]{lieb88}, full circles; \cite[Evans (1992)]{evan92}, full squares; 
\cite[Oswalt \etal, (1996)]{oswa96}, open triangles; \cite[Legget \etal, (1998)]{legg98},
open diamonds; \cite[Knox, Hawkins \& Hambly (1999)]{knox99}, open circles.}} 
\end{minipage}\hspace{2pc}%
\begin{minipage}{15pc}
\includegraphics[width=15pc]{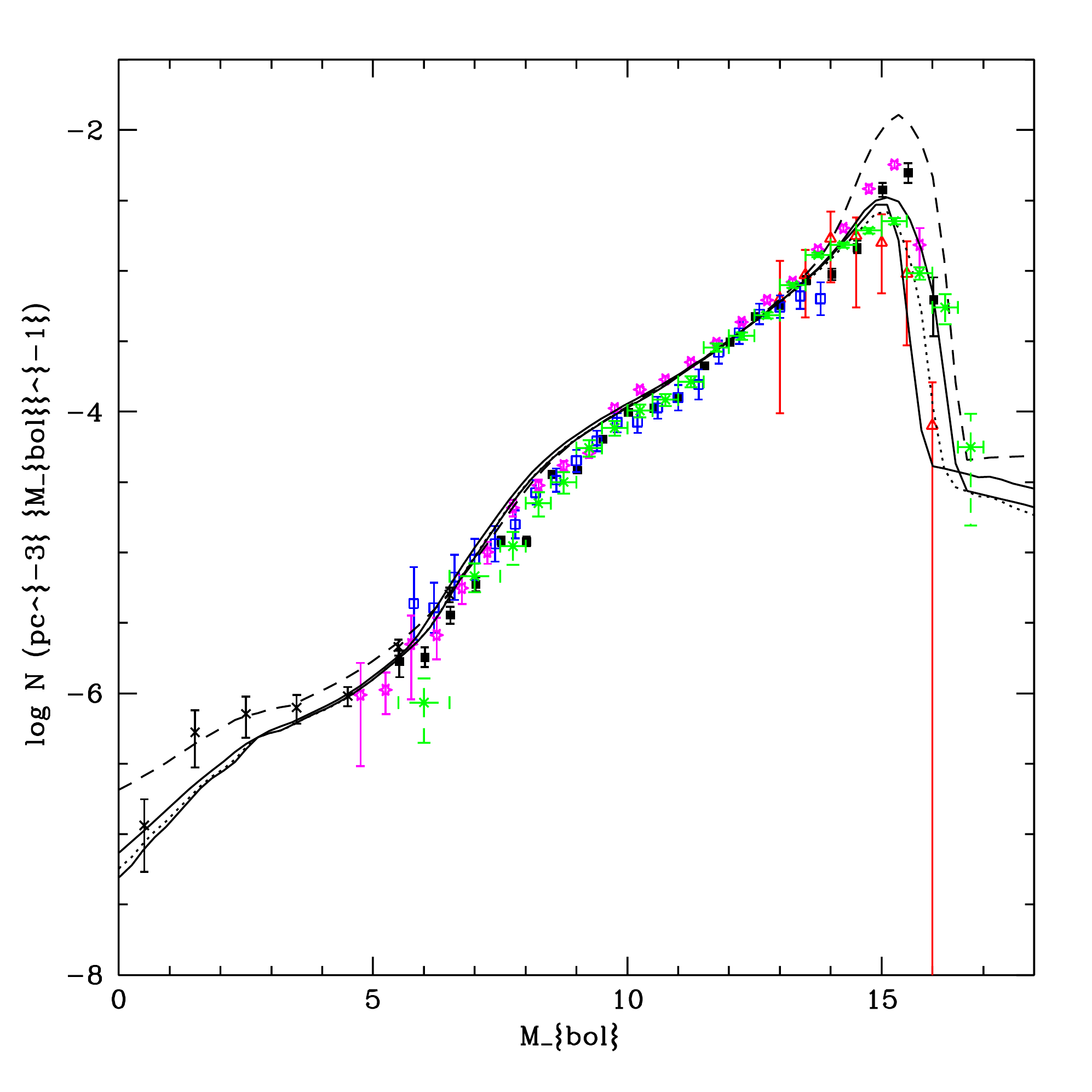}
\caption{\label{fig4} {\footnotesize Luminosity functions obtained from the SDSS 
catalogue: Black solid squares (\cite[Harris \etal, 2006]{harr06}), open blue squares, only DAs (\cite[DeGennaro \etal, 2008]{dege08}), black crosses  (\cite[Krzesinski \etal, 2009]{krze09}), and green stars (\cite[Munn \etal, 2017]{munn17}). Magenta stars were obtained from the SCSS catalogue (\cite[Rowell \& Hambly (2011)]{rowe11}) and contain DA and non--DA stars. }}
\end{minipage} 
\end{figure}

The first empirical luminosity function was obtained by \cite[Weidemann (1968)]{weid68} and was improved by several authors during ninetees (Figure~\ref{fig3}) proving in this way  that the evolution of white dwarfs was just a cooling process and that there was a cut-off in the distribution caused by the finite age of the Galaxy. The position of the cut-off is sensitive to the cooling rate and, consequently, it 
can be used to constrain any new theory  or hypothesis implying the introduction of an additional 
source or sink of energy. However the low number of stars in the samples, few hundreds, and 
the uncertainties in the position of the cut-off prevented anything else than obtaining upper 
bounds.

The advent of large cosmological surveys like the Sloan Digital Sky Survey (SDSS) and the Super COSMOS Sky Survey, both completely independent, introduced a noticeable improvement in the precision and accuracy of the luminosity function since they allowed to increase the sample size to several thousands of stars. Figure~\ref{fig4} displays both functions normalized to the  $M_{\rm bol} \approx 12$ bin. As can be seen, they nearly coincide over a large part of the brightness interval. At large brightness, $M_{\rm bol} < 6$, both luminosity functions show a large dispersion, not plotted in the figure, as a consequence of the fact that the proper motion method is not appropriate there. One way to circumvent this problem relies on the UV-excess technique (\cite[Krzsinski et al. 2009]{krze09}). The results obtained in this way are represented by black crosses  after matching their dim region with the corresponding bright segment  of the \cite[Harris et al. (2006)]{harr06} distribution. As a complement, the luminosity function of the dimmest white dwarfs obtained by \cite[Legget et al. (1998)]{legg98} has been included (red triangles).  The discrepancies at low luminosities are due to the difficulty to separate DAs from non-DAs and to the different behavior of the envelope. 

The quality of this new luminosity functions allowed, for the first time, to determine their shape and to use the slope as a tool to test new physical theories.  If  an additional  source or  sink  of energy  is  added, the  characteristic cooling time  is modified  and its imprint  appears in  the luminosity function, as can  be seen  in Fig.~\ref{fig4},  where a  change of slope is evident at magnitudes $M_{\rm bol}\sim 8$. This change is caused by the transition from a cooling dominated by neutrinos to one dominated by photons. As an example, this technique was used by \cite[Isern \etal, (2008)]{iser08b} to suggest that axions of the DFSZ type could be contributing to the cooling of white dwarfs.

The main problem when using Eq.~\ref{lf} is that the star formation rate has to be obtained independently from white dwarfs in order to break the degeneracy between the galactic properties and stellar models. Fortunately, the luminosity function has an important property. The shape of the bright branch, stars  brighter than $M_{bol} \approx 13$, is almost independent of the assumed star formation rate as it can be seen in Fig.~\ref{fig4} where several theoretical white  dwarf luminosity functions are displayed. The two solid black lines have been obtained assuming  a  constant SFR but two ages of  the Galaxy, 10 and 13~Gyr respectively, the dashed line assuming a decreasing exponential SFR, $\Psi \propto \exp (-t/\tau),\, \tau= 3\,{\rm Gyr} $, where $t$ is the age of the Galaxy, and the dotted line assuming an almost constant SFR with an exponentially decreasing tail that represents models where the star formation propagated from the center to the periphery,
 $\Psi \propto (1+\exp [(t-t_0)/\tau])^{-1}, \tau = 3 \, {\rm Gyr},\, t_0= 10\,{\rm Gyr}$, where $t$ is the looking back time. 
As can be seen, in the region $  M_{\rm bol}\approx 6 -13$, all luminosity 
functions overlap as far as the SFR is smooth enough.  The differences due to the shape of  the SFR only appear  in the regions containing  cool or very
bright  white   dwarfs.   Unfortunately,  the  observational
uncertainties in these regions prevent at present to discriminate among the different possibilities.

This behavior can be understood in the following way: Eq.~(\ref{lf}) can be written as:

\begin{equation}
n\left( l \right) \propto \left\langle {{\tau _{cool}}} \right\rangle \int\limits_{{M_i}}^{{M_{\max }}} {\Phi \left( M \right)\Psi \left( T_G-t_{cool}-t_{ps}  \right)} dM
\label{lftau}
\end{equation}
Restricting  this equation to the bright white  dwarfs --  namely, those  for
which $t_{\rm  cool}$ is small --  the lower limit of  the integral is
satisfied  by low-mass  stars  and,  as a  consequence  of the  strong
dependence of the main sequence lifetimes with mass, it adopts a value
that  is almost  independent  of the  luminosity under  consideration.
Therefore,  if $\Psi$  is a  well-behaved  function and  $T_G$ is  large
enough,  the lower limit is almost independent of the luminosity, 
and the value of the integral is incorporated  into  the normalization  
constant  in such a way that  the shape  of  the
luminosity function  only depends on the  averaged physical properties
of the white dwarfs (\cite[Isern \& Garc\'{\i}a-Berro, 2008]{iser08}).  This average is dominated by low mass white dwarfs and, as far as the mass spectrum is not strongly perturbed by the adopted star formation or the initial mass function, it is representative of the intrinsic properties of white dwarfs (\cite[Isern \etal, 2009]{iser09}). This shape, however, can be modified by a recent burst of star formation since, in this case, low-mass main sequence stars have no time to become white dwarfs and $M_I$  in Eq.~\ref{lftau} becomes luminosity dependent. On the contrary, if the burst is old enough, the corresponding luminosity functions are barely modified.

Another important property is that in the bright region considered here, the slope of the relationship between the luminosity and the core temperature of DA and non--DA white dwarfs almost coincide,  and both luminosity functions almost overlap in this luminosity interval after normalization. This is the reason why the luminosity function of \cite[DeGennaro \etal, (2008)]{dege08} containing only DAs coincidesafter normalization with those containing DAs and non--DAs as it can be seen in Fig.~\ref{fig4}.
Therefore, Eq.~\ref{lftau} offers the possibility to use the slope of the bright branch of the luminosity function to detect the presence of unexpected additional sinks or sources of energy in white dwarfs. In the case of axions, this method was used for the first time by \cite[Isern \etal. (2008)]{iser08a} who obtained $g_{ae} =(1.4^{+0.9}_{-0.8})\times 10^{-13}$, see WDLF08 label in Fig.~\ref{fig2}, using a preliminary version of the Harris \textit{loc cit.} \cite[Miller Bertolami et al. (2014)]{mill14} reexamined this result using all the luminosity functions of Fig.~\ref{fig4} (but with a consolidated Harris \textit{loc cit.} luminosity function) and with a self consistent treatment of the neutrino cooling and concluded that DFSZ axions with a coupling constant $g_{ae}$ in the range of  $(0.7-2.1)\times 10^{-13}$ could exist (WDLF14 label in Fig.\ref{fig2}). All these results, however, have to be regarded as qualitative, since the uncertainties plaguing the determination of both, observed and computed, luminosity functions are still large.

Given the degeneracy between the stellar and galactic terms in  Eq.~\ref{lftau}, it is natural to wonder if the changes in the shape of the luminosity function attributed to axions are an artifact introduced by the star formation rate. One way to break this degeneracy and decide if it is necessary to introduce or not new physical effects is to examine the luminosity function of populations that have different formation histories. If axions exist and can modify the cooling of white dwarfs, their imprint will be present in all the luminosity functions at roughly the same luminosities.
Furthermore, it is well known that the adopted white dwarf scale height above the galactic plane s has a noticeable effect on the shape of the luminosity function (\cite[Garc\'{\i}a-Berro \etal, 1988]{garcb88};\cite[Harris \etal, 2006]{harr06};\cite[Kilic \etal, 1917]{kili17}). This argument reinforces the convenience of analyzing the luminosity function of white dwarfs with different scale heights and independent star formation histories (\cite[Isern \etal, 2018]{iser18}).

\begin{figure}[h]
\center
\includegraphics[width=1.0\textwidth, clip=true, trim= 2cm  4cm 0cm 7cm]{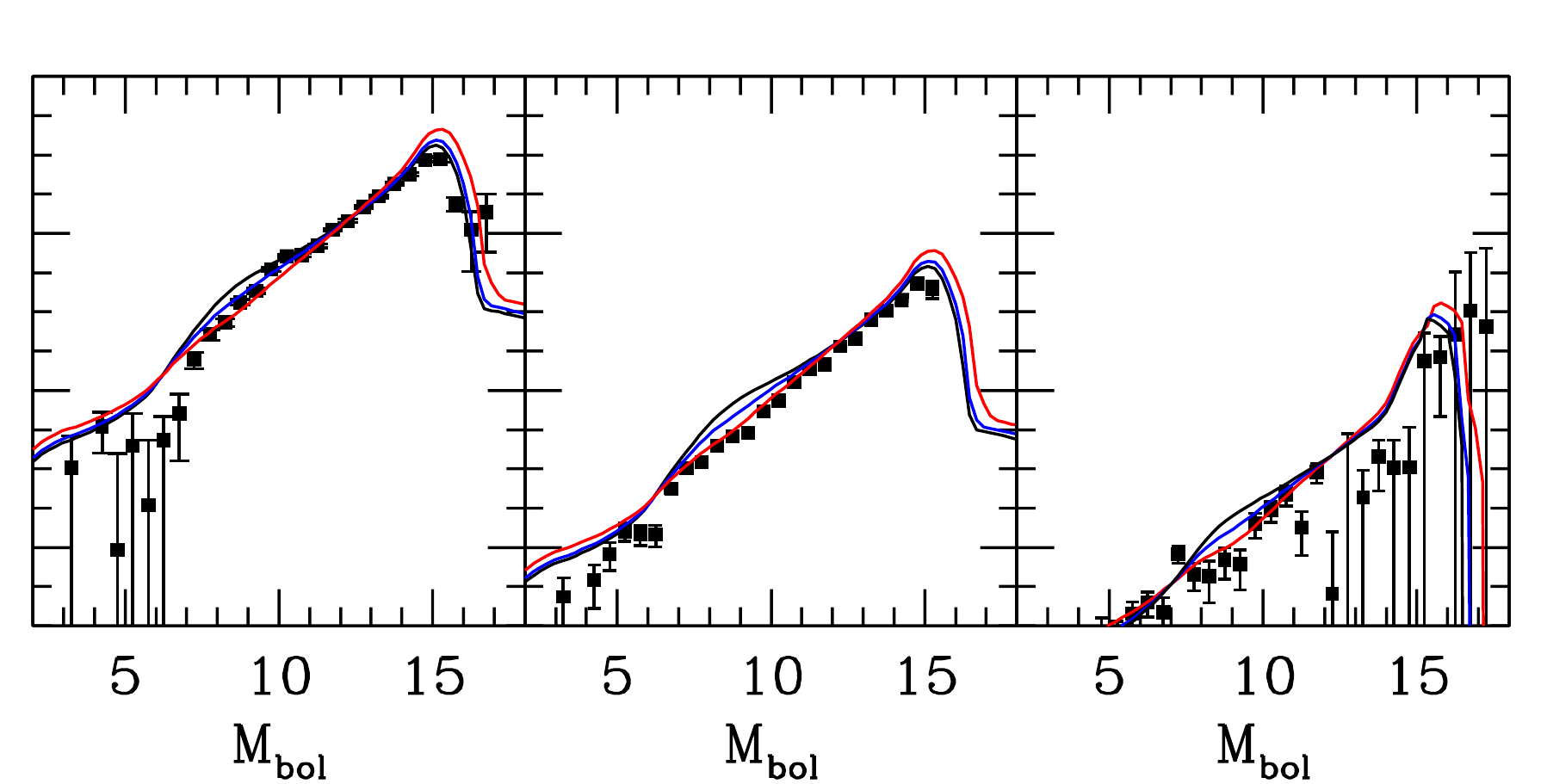}
  \caption{\footnotesize Luminosity functions of DA and non-DA white dwarfs of the thin and thick discs and halo [\cite[Rowell \& Hambley 2011]{rowe11}]. The black line represent the case without axions. The blue and red lines correspond to the cases where DFSZ axions with coupling constants $g_{ae}\times 10^{13}= 2.24$ and $4.48$ are included in the cooling model .}
\label{fig5}
\end{figure}

\cite[Rowell \& Hambly (2011)]{rowe11} provided, for the first time, a luminosity function for the thin and thick disks  and noticeably improved  that of the halo (Fig.~\ref{fig5}). Table~\ref{tab1} shows how the discrepancies between theoretical calculations and observations decrease in all three cases when axions are included. The fit favors axions within the interval $g_{ae}=(2.24 -4.48) \times 10^{-13}$ with some tension between the halo and the disk results, although the $2\,\sigma$ bounds are compatible. See Fig.~\ref{fig2}, label WDLF(RH). 

\begin{table}
	\centering
	\caption{Reduced $\chi^2$ obtained from the fitting of the observed white dwarf luminosity functions in the brightness interval $6\, < \, M_{\rm bol} \, < \, 12.5$ with different intensities of the coupling between electrons and axions [Isern et al. 2018].}
	\label{tab1}
	\begin{tabular}{lcccc} 
		\hline
		$g_{\rm ae}\times 10^{13}$   & 0.00 & 1.12 & 2.24 & 4.48 \\
                 $m_{\rm a} \cos^2 \beta$ (meV) & 0  &     4   &   8    & 16    \\
		\hline
                   Rowell \& Hambly (2011)     &         &         &         &          \\
                   \hline
                   thin disk                             &18.59&15.33& 7.33 &21.52 \\
                   thick disk                           &30.72&26.61& 11.73& 1.43 \\
                   halo                                    &  3.31& 2.94 &  2.36 & 1.87  \\
                   \hline
                   Munn et al. (2017)                &          &         &         &          \\
                   \hline
                   thin+thick disk                  &  4.89 & 2.87 &  1.46 &  6.13 \\
                   halo                                    &  2.46 & 1.50 &  0.65 & 1.42 \\
		\hline
                   Kilic et al. (2017)                  &          &         &         &          \\
                   \hline
                    $\Phi_{200-900}$             &  3.25 & 2.09 &  1.05&  6.11  \\
                    $\Phi_{200-700}$             &  3.53 & 2.32 &  1.14&  5.80  \\
                    $\Phi_{200-500}$             &  4.02& 2.73  &  1.28&  2.26  \\
                    $\Phi_{300}$                     &  6.11& 4.95  &  2.26&   4.98 \\
		\hline
	\end{tabular}
\end{table}

\begin{figure}[h]
\begin{minipage}{15pc}
\includegraphics[width=15pc,clip=true, trim=0cm 0cm 1cm 4.5cm]{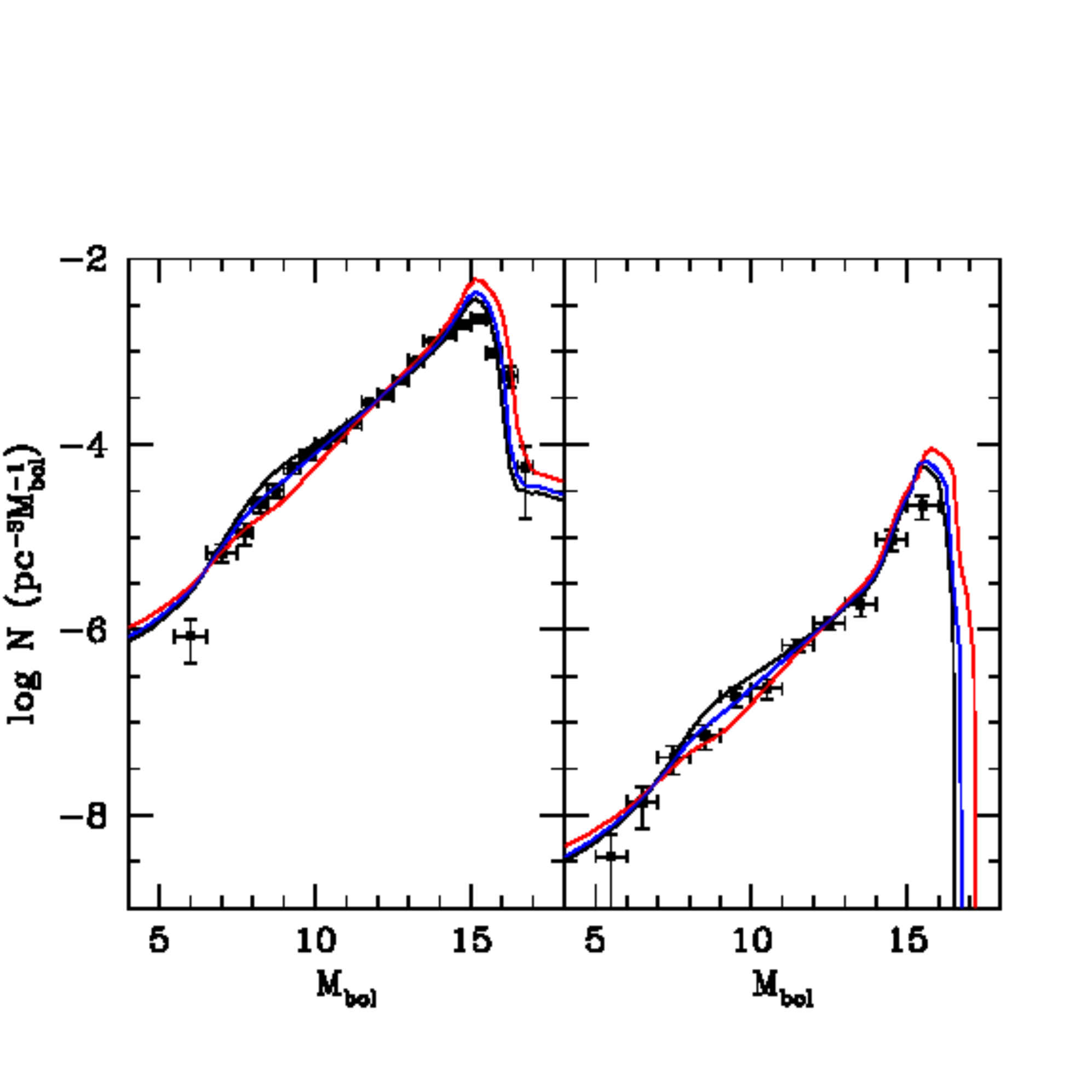}
\caption{\label{fig6} {\footnotesize Luminosity functions  of DA and nDA white dwarfs belonging to  the disk (thin and thick) and the halo (Munn \textit{et al.}, 2017). The meaning of the solid lines is the same as in Fig.~\ref{fig5}.} }
\end{minipage}\hspace{2pc}%
\begin{minipage}{15pc}
\includegraphics[width=15pc,clip=true, trim=0cm 6cm 0cm 3.5cm]{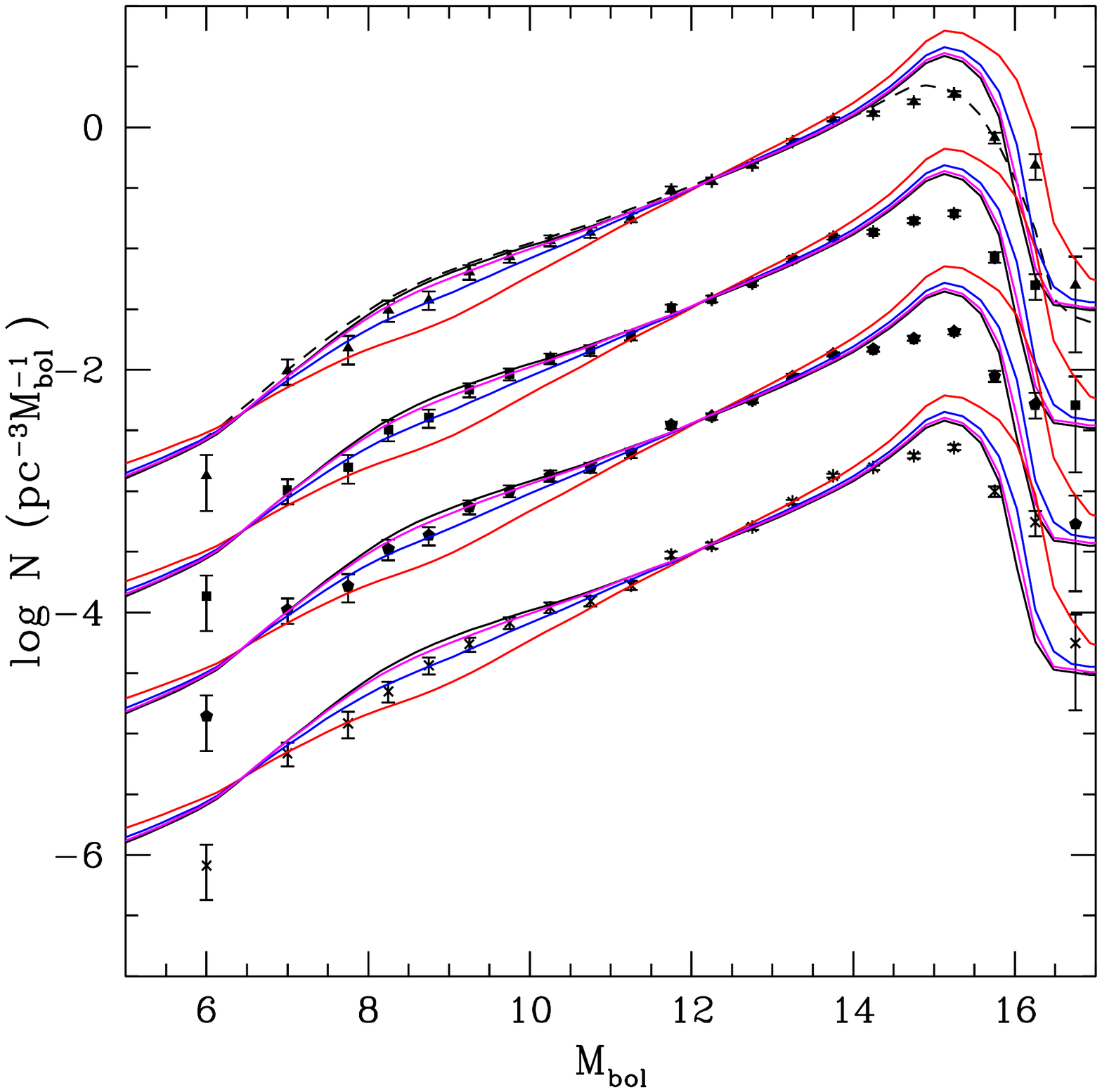}
\caption{\label{fig7} {\footnotesize White dwarf luminosity functions of the disk assuming different scale heights as proposed by Kilic \textit{et al.} (2017). From top to bottom $\Phi_{200-900} +3$, $\Phi_{200-700} +2$,  $\Phi_{200-500} +1$ and $\Phi_{300}$. Solid lines correspond to $g_{ae}\times 10^{13}=0, 1.12, 2.24, 4.48$ (black, magenta, blue, red, respectively). The dashed line corresponds to a case with no axions and an SFR constant plus an exponential tail as in Fig~\ref{fig4}.}}
\end{minipage} 
\end{figure}

Munn \textit{et al.} (2017) improved the Harris\textit{ et al.} (2006) luminosity function resolving the peak and increasing the precision of the brighter branch assuming a constant scale height above the Galactic plane. They also computed the luminosity function for the halo selecting white dwarfs with $ 200 \le v_{\rm tan} \le 500$~km~s$^{-1}$. As before, the inclusion of axions improves the concordance between theory and observations both in the disk and the halo. The best fit is obtained for $ g_{ae} \approx 2.24 \times 10^{-13}$, label WDLF(M) Fig.~\ref{fig2}, with ($2\,\sigma$) upper bounds of $g_{ae} <(4.2\, {\rm and}\, 14) \times 10^{-13}$ coming, respectively, from the disk and halo luminosity functions.

Kilic \textit{et al.} (2017) reexamined the Munn \textit{et al.} (2017) data assuming three variable scale heights above the galactic plane going linearly from 200 pc now to 900, 700 and 500 pc in the past ,respectively, and one with a fixed scale height of 300 pc. They concluded that the slight discrepancies in the region $6 \le M_{\rm bol} \le 12.5$ were caused by the use of a fixed scale height. This argument is correct but, as it was shown by Isern et al (2018), see Fig.~\ref{fig7}, the inclusion of axions improves the agreement. In these cases the best fit was also around $g_{ae} \approx 2.24 \times 10^{-13}$, Table~\ref{tab1}, and a $2\,\sigma$ global upper bound $g_{ae} < 6 \times 10^{-13}$ (Fig.~\ref{fig2}, label WDLF(K)).

\begin{figure}[h]
\center
\includegraphics[width=1.0\textwidth, clip=true, trim= 0cm  3cm 0cm 0.5cm]{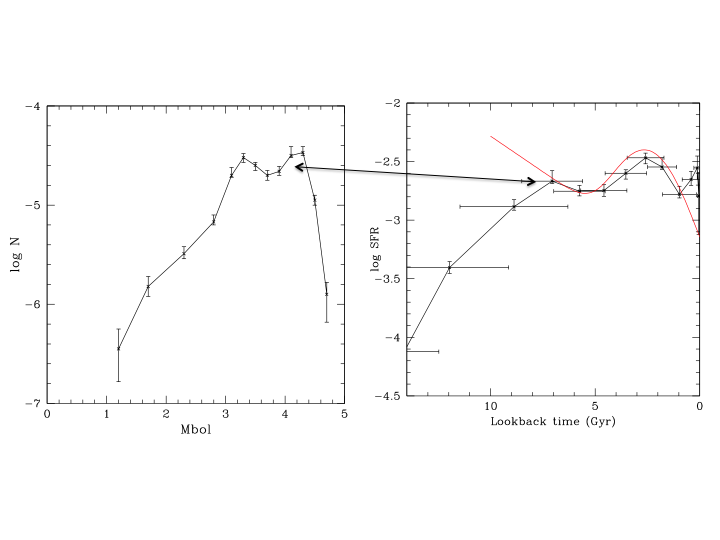}
  \caption{\footnotesize Left panel: empirical luminosity function of massive ($ M_{\rm WD} \le M/M_\odot \le 1.1 $) DA white dwarfs in the solar neighborhood  (\cite[Tremblay \etal, 2019]{trem19}). Right panel: in black, the corresponding star formation rate (\cite[Isern, 2019]{iser19}), in red, the Galactic disc star formation (\cite[Mor \etal, 2019]{mor19}.}
\label{fig8}
\end{figure}

Recently a second possibility has appeared.  \cite[Tremblay \etal,  (2019)]{trem19} have been able to build a reliable and precise luminosity function of massive white dwarfs that belong to the solar neighborhood ($ d\le 100$~pc) using the data provided by Gaia (Fig.~\ref{fig8}, left panel). 
If the luminosity function, Eq.~\ref{lf}, is restricted to massive white dwarfs, i.e. those for which it is possible to neglect the lifetime of the progenitor  in front of the cooling time, and $\Psi$ is smooth enough,  the age of any bin and the star formation rate corresponding to this time can be computed as

\begin{equation}
\left\langle t \right\rangle  = \frac{{\int\limits_{\Delta M} {\Phi (M)tdM} }}{{\int\limits_{\Delta M} {\Phi (M)dM} }}
\end{equation}
and
\begin{equation}
\left\langle \Psi  \right\rangle  = \frac{{n(l)}\Delta l}{{\int\limits_{\Delta M} {\Phi \left( M \right)\Delta {t_{cool}}\left( {l,M} \right)dM} }}
\label{psi}
\end{equation}
with
$
\Delta {t_{cool}} = {t_{cool}}\left( {l + 0.5\Delta l,M} \right) - {t_{cool}}\left( {l - 0.5\Delta l,M} \right)
$,
$l=-\log(L/L_\odot)$ and $\Delta l$ the width of the luminosity function bin (Isern, 2019) and this allows to reconstruct the star formation history of the solar neighborhood\footnote{It is important to notice here that the star formation rate obtained in this way  does not take into account the secular evolution of the sample caused by radial migrations, scale height  inflation or merging of double degenerate stars.} (Fig.~\ref{fig8}, right panel).  The star formation rate obtained in this way  is not constant or monotonically decreasing as it is often assumed. It grew quickly in the past, during the first epochs of the Galaxy, it roughly stabilized and started to decrease 7 to  6 Gyr ago. A noticeable feature is a prominent peak centered around 2.5 Gyr ago, the exact position being model depending.

\begin{figure}[h]
\begin{minipage}{15pc}
\includegraphics[width=15pc,clip=true, trim= 0cm  4cm 0cm 0.5cm]{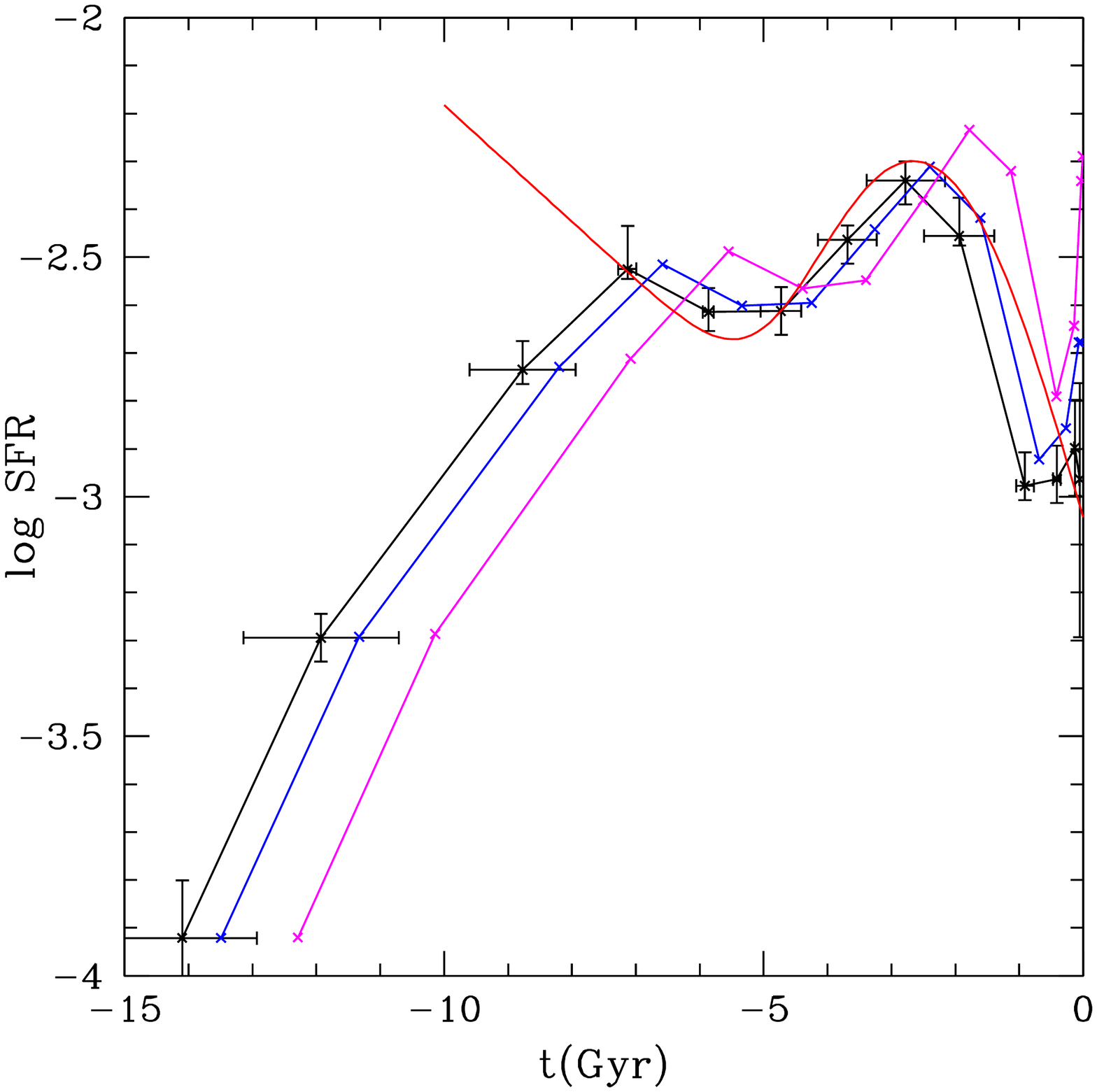}
\caption{\label{fig9} {\footnotesize Star formation rate (M$_\odot$Gyr$^{-1}$pc$^{-3}$) in the solar neighborhood obtained from massive white dwarfs assuming no axions (black line). The blue and magenta lines correspond to the case  when axions are present and have a coupling constant $g_{ae}=2.24\,{\rm and}\,4.48\times 10^{-13}$, respectively. The red line corresponds to the star formation rate obtained by \cite[Mor \etal, (2019)]{mor19}.}}
\end{minipage}\hspace{2pc}%
\begin{minipage}{15pc}
\includegraphics[width=15pc,clip=true, trim= 0cm  4cm 0cm 3.5cm]{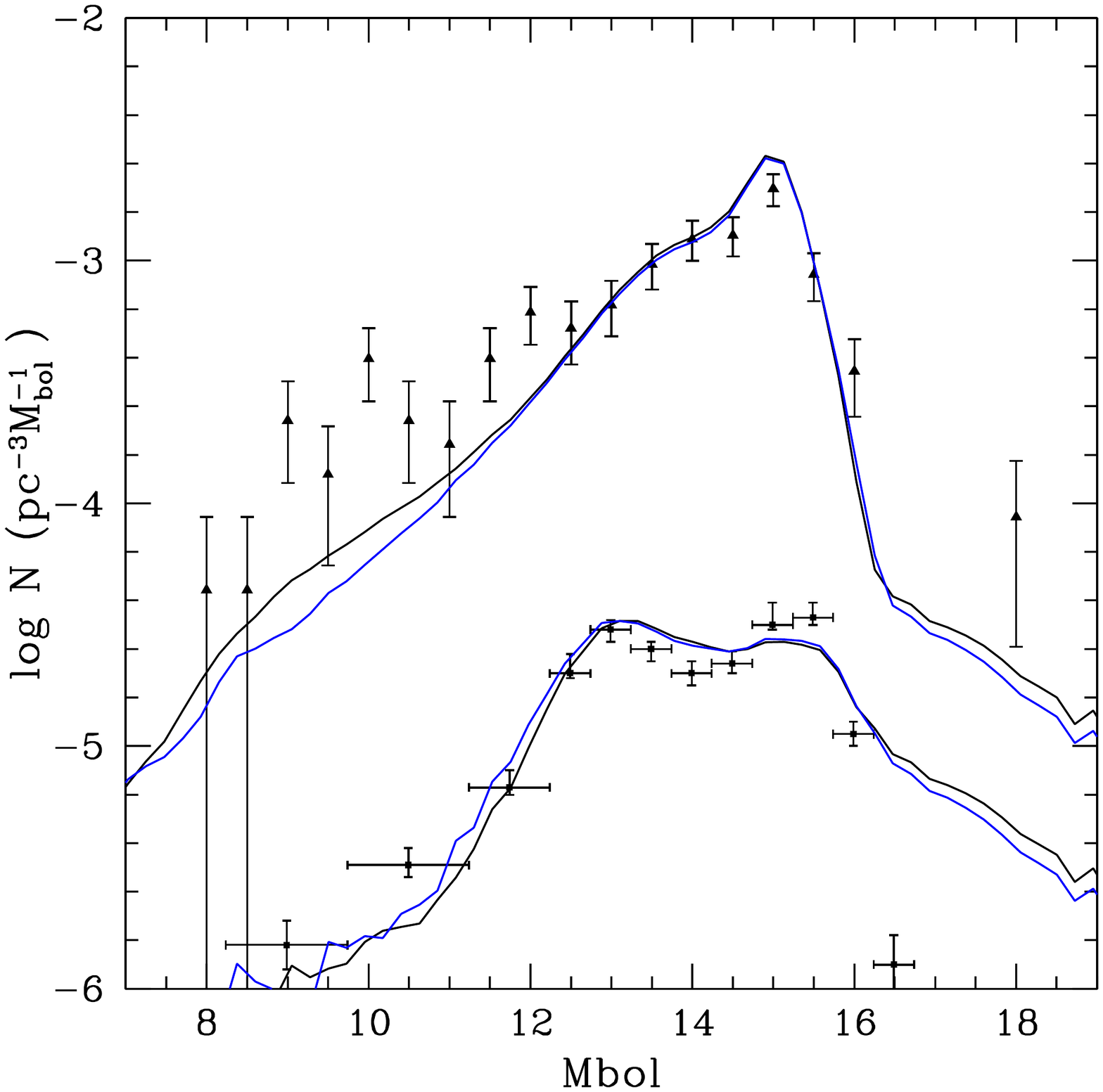}
\caption{\label{fig10} {\footnotesize Theoretical versus empirical luminosity functions for massive and all white dwarfs. The empirical ones are from \cite[Tremblay \etal (2019)]{trem19} and \cite[Oswalt \etal (2017)]{oswa17} respectively.The theoretical  ones were  obtained  with the star formation rates of Fig.~\ref{fig9} with $g_{ae}=0$ (black lines) and $2.24\times 10^{-13}$ (blue lines).}}
\end{minipage} 
\end{figure}

The existing degeneracy between galactic properties and evolutionary models imply that different stellar models can lead to different star formation histories, for which reason it is necessary to compare these results with others obtained independently. \cite[Mor \etal, (2018)]{mor18} computed the star formation history of the Galactic disk using main sequence stars from the Gaia DR2 and the Besan\c con Galaxy Model. Since this function is expressed in stars per unit of disk surface it has been divided by an arbitrary and constant height scale above the galactic plane (red line, Fig.~\ref{fig9}). As it can be seen both methods, local and galactic,  predict a concordant burst of star formation $\sim 2.5$~Gyr ago but diverge at early times. This divergence may have several origins, a local delay in starting the star formation process, a different behavior of the outer and inner disks, a vertical dilution caused by a galaxy collision or just the conversion of DA white dwarfs into non-DAs. The peak that appears at $\sim 0.2$~Gyr is in the limit of applicability of the method and deserves more attention.

Figure~\ref{fig9} displays the star formation rate obtained when including axions (and crystallization sedimentation effects) in the cooling models. As it can be seen, the SFRs obtained in this way have the same behavior, but they are displaced towards shorter ages and larger rates thus providing an additional way to discriminate among theoretical models. Figure~\ref{fig10} displays a test of consistency where the SFR obtained from the luminosity function of the massive ones is used to compute the total one. The agreement between theory and observations is reasonable taking into account that observations were obtained independently and that the results are very sensitive to the DA/nDA properties as well as to the adopted initial mass function and initial final mass relationship. The excesses around $M_{\rm bol} \sim 9-10$ could be due to recent bursts of star formation not resolved by the present binning of the massive luminosity function and, as mentioned before, deserves a detailed analysis. Figure~\ref{fig9} also shows that values of $g_{ae} \gapprox 3\times 10^{-13}$ are not compatible with the results obtained by Mor \textit{et al.}, (2019). 

\section{Discussion and conclusions}
The two methods presented here are complementary since they measure the cooling rate of the same object using different phenomenologies. One, the drift of the pulsation period, applies to individual stars, while the other, the luminosity function, applies to the population ensemble. As it is evident from Fig.~\ref{fig2}, there is some tension between both sets of data but, given the uncertainties, they can be considered as qualitatively concordant.

Axions with the properties described here not only would perturb the cooling of white dwarfs but they would also modify in a subtle way the evolution of other kind of stars. An extensive review can be found in \cite[Giannotti \etal, (2017)]{giann17}, while \cite[Sedrakian (2019)]{sedr19} provides a recent analysis of the cooling by DFSZ axions of the neutron star in Cas A. Here only the two more restrictive tests are presented.

The luminosity of a star in the red giant branch depends on the mass of its core and is due to the hydrogen burning in a thin shell surrounding it (\cite[Paczynski, 1970]{pacz70}). Thus, the luminosity grows with the core until He is ignited in the center. If axions were present, the core would be more massive than in the standard case and, consequently, the tip of the red giant branch would be brighter. Obviously, since H-burning occurs via CNO-cycle, this luminosity depends on the metallicity. The analysis of the red tip in M5 suggested the necessity of an extra cooling source. If this source were axions, the electron coupling constant should be $g_{ae} \sim 1.9 \times 10^{-13}$ (\cite[Viaux \etal, 2013]{viau13}), but a similar analysis performed on M3 did not find the necessity of an extra cooling term and provided a more stringent upper bound of $g_{ae}<2.57\times 10^{-13}$ (\cite[Straniero \etal, 2018]{stra18}). A similar bound has also been obtained using a set of 50 globular clusters (\cite[Arceo-Diaz \etal, 2019]{arce19}). In Fig.~\ref{fig2} this bound and the  hint found in M5 are represented by the RGB-T line. A detailed analysis of the existing uncertainties can be found in \cite[Serenelli \etal, (2017)]{sere17}. 

The evolution of HB stars provides additional observational tests.  The number of stars in a given region of the HR diagram is roughly proportional to the time spent in it. Since HB stars are the direct descendants of red giant stars, the ratio between the number of HB stars, $N_{\rm HB}$, and the red giants, $N_{\rm RGB}$,  in a cluster should satisfy the relationship $R=N_{\rm HB}/N_{\rm  RGB}=\Delta t_{\rm HB}/\Delta_{\rm  RGB}$ and, since the densities and temperatures of both populations  are very different, the inclusion of the axion emission should strongly perturb these quantities. This parameter has been measured in a large sample of globular clusters and is fairly constant at low metallicities, $R_{\rm av} =1.39 \pm 0.03$ (\cite[Salaris \etal, 2004]{sala04}). 

\cite[Ayala \etal, (2014)]{ayal14} and \cite[Straniero \etal,  (2015)]{stra15}  examined the influence of axions in this case and found that the best fit to the parameter R was obtained for $g_{a\gamma} =0.29\pm 0.18\times 10^{-10}$~GeV$^{-1}$, i.e. that it was necessary an extra cooling factor, with an upper bound of $g_{a\gamma} <0.66 \times 10^{-10}$~GeV$^{-1}$. In terms of the mass of the axion, these values translate to $0.12$ eV and $<0.2$ eV, respectively, in the case of the KSVZ axions. Since they did not include bremsstrahlung with electrons, that strongly affects the evolution of RGB stars, it is not possible to extrapolate these values to the DFSZ case. However, in an updated version, taking into account the axion-electron interaction, \cite[Straniero \etal, (2018)]{stra18} obtained  $g_{a\gamma} <0.5 \times 10^{-10}$~GeV$^{-1}$ and $g_{ae} <2.6 \times 10^{-13}$ (these values are represented in Fig.~\ref{fig2}). These results, however, strongly depend  on the assumed He abundance in the cluster.

The \emph{Zero Age Horizontal Branch} provides an additional test. If the axion-electron interaction is operating in RGB stars, the core is larger and the luminosity of HB stars too. \cite[Straniero \etal, (2018)]{stra18} examined this case and found a best fit at $g_{ae} =0.54 \times 10^{-13}$ and a bound at $g_{ae} <1.78 \times 10^{-13}$. These values are plotted in Fig.~\ref{fig2}.

Certainly, the deviations represented in Fig.~\ref{fig2} have a small statistical significance, but taken together they suggest that stars loose energy more efficiently than expected. Axions of the DFSZ type, with a mass of few meV could do the job, although other possibilities are open. For instance an ALP\footnote{Axion-like particle (ALPs) appear in a natural way in the extensions of the Standard Model as pseudo Nambu-Goldstone bosons associated to the breaking of symmetries (\cite[Ringwald, 2014]{ring14}). They are similar to axions but mass and photon coupling are no longer linked, thus offering a larger space of parameters.} with $g_{ae}\sim  1.5\times 10^{-13}$ and $g_{a\gamma}\sim 1.4\times 10^{-11}$~GeV$^{-1}$ (\cite[Giannotti \etal, 2016]{giann16}). 
It is evident from this discussion that a direct detection is necessary to prove the existence of axions and that white dwarfs provide an interesting hint on where to search. IAXO is a future experiment designed to detect the emission of axions  by the Sun (\cite[Irastorza \etal, 2011]{iras11}) that will have enough sensitivity to detect axions of the DFSZ type and mass  $\gapprox 3$~meV, according to the calculations of \cite[Redondo, (2013)]{redo13}.

Thanks to the data that is providing Gaia and will provide the LSST as well as other instruments, white dwarfs will become in a next future one of the best characterized populations of the Galaxy, and this will convert them into an excellent laboratory to explore new ideas in Physics.

\section*{Acknowledgements}
This work  has been supported  by MINECO grant  ESP2017-82674-R, 
by EU FEDER funds, and by
grants 2014SGR1458 and CERCA Programe of  the Generalitat de
Catalunya.

\end{document}